\documentclass[10pt,twoside]{article}
\usepackage[english]{babel}

\usepackage {graphicx}
\usepackage {graphics}
\usepackage {floatflt}
\usepackage {latexsym}
\usepackage {caption3} 
\usepackage[labelsep=period]{caption}[2007/11/04 v.3.1e]
\usepackage {amssymb}

\begin{document}

\begin{center}
\sl\small{Proceedings of XV International Scientific Meeting "Physical Interpretations of Relativity Theory" \\
\hspace {-52mm} (PIRT-2009, Moscow 6-9 July), Moscow-Liverpool-Sunderland.} \\
\smallskip
\hspace {-68mm}{\underline{\scriptsize{{{\sl\small{\copyright{Bauman Moscow State Technical Unversity, 2009}}}}}}} 
\smallskip
\end{center}

\vspace{1cm}

{\large\bf CHARGED BALL STATIC STAR MODEL \footnote{\small{Talk presented at Proceedings of XV International Scientific Meeting "Physical Interpretations of Relativity Theory" (PIRT-2009), Moscow 6-9 July, 2009.}}}

\bigskip

{\large\bf Baranov A.M., Vlasov Z.V.}

Siberian Federal University

Department of Theoretical Physics

Russia, 660041, Krasnoyarsk, Svobodny Av., 79

E-mail:  alex\_m\_bar@mail.ru

\bigskip

Within the framework of General Relativity the model approach to a description of spherical gravitating static fluid balls with an electric charge is considered. The metric interval is written in Bondi's radiation coordinates. The total energy-momentum tensor as a direct sum of the perfect Pascal fluid energy-momentum tensor and the energy-momentum tensor of an electromagnetic field is chosen. The exact solution of the Einstein-Maxwell equations as an extension of similar solution with parabolic distribution of mass density is found.


\section{Introduction}

The problem of exact solutions finding of Einstein's equations within of the General Relativity does not lose a significance to this day. Special interest, despite of an exoticism, represents models of electric charged stars because an electrical charge is one of the few physical quantities which not disappears at a star blow-up and the star collapse. The basic difficulties of a calculation of similar models are concluded in nonlinearity of the combined equations of Einstein. It is main insufficiency for system description. Other difficulty is related to the solution of self-consistent  of Einstein-Maxwell combined equations. The similar approach is realised earlier by one of authors in [1] where the exact solution for the charged Pascal perfect fluid with parabolic distribution of mass density  for a neutral component of fluid is discovered, but without an introduction of a concrete equation of state. 

The generalisation of this model related to a modification of the mass density distribution law of a neutral fluid is considered in given article. Such a modification leads to a new distribution of the charged density into a ball model and to a new exact static spherical solution of Einstein-Maxwell combined equations. In this case as usually the exterior gravitational field should be described by Reissner-Nordstr$\ddot{o}$m's solution for the charged spherical mass as the generalisation of exterior Swarzschild's solution.

We make choose a static and spherical model here. The metric functions do not depend on a time variable. Such a model possesses spherical symmetry that removes dependence on angular variables. The model is considered without a rotation and a radiation.

\section{The basic mathematical expressions}

\indent 

Here we write a metric in Bondi's radiation coordinates

\begin{equation} 
ds^2 = F(r)dt^2 +2L(r)dtdr-r^2(d{\theta}^2 + sin{\theta}^2d{\varphi}^2),
\label{eq:1}
\end{equation}

\noindent 
where $F=F(r)$ и $L=L(r)$ are the metric functions of a radial variable $r$; $\; t$ is a time coordinate; $\;\theta$ and $\varphi$ are angle variables; the light velocity is chosen equal to unit as a gravitational constant of Newton $G_{N}=1$. A determinant of a covariant metric tensor $g_{\alpha\beta}$ corresponding to an expression (\ref{eq:1}) is $det(g_{\alpha\beta})\equiv g=-L^{2}r^{4}sin^{2}\theta.$
 
We will set tetrads or an orthonormal 4-basis in a tangential space-time with using the metric (\ref{eq:1}) as 

\begin{equation} 
g_{(0)\mu}=\delta^{0}_{\mu}; \quad g_{(1)\mu}=L\delta^{1}_{\mu}+\frac{1}{2}F\;\delta^{0}_{\mu}; 
\label{eq:2}
\end{equation}

\begin{equation} 
g_{(2)\mu}=-\frac{r}{\sqrt{2}}(\delta^{2}_{\mu}+i\, sin\theta\;\delta^{3}_{\mu}); \quad g_{(3)\mu}=-\frac{r}{\sqrt{2}}(\delta^{2}_{\mu}-i\, sin\theta\;\delta^{3}_{\mu}); 
\label{eq:3}
\end{equation}

\begin{equation} 
g^{\mu}_{(0)}=L^{-1}\;\delta^{\mu}_{1}; \quad g^{\mu}_{(1)}=\delta^{\mu}_{0}+\frac{1}{2}FL^{-1}\;\delta^{\mu}_{1};
\label{eq:4}
\end{equation}

\begin{equation} 
g^{\mu}_{(2)}=\frac{1}{r\sqrt{2}}(\delta^{\mu}_{2}+\frac{i}{sin\theta}\;\delta^{\mu}_{3}); \quad g^{\mu}_{(3)}=\frac{1}{r\sqrt{2}}(\delta^{\mu}_{2}-\frac{i}{sin\theta}\;\delta^{\mu}_{3}),
\label{eq:5}
\end{equation}

\noindent
where $i$ is the imaginary unit and Greek indices run  through $0, 1, 2, 3$.

 Now we will determine basic differential 1-forms of Cartan with helping of the tetrads (\ref{eq:1})-(\ref{eq:5}) as contractions of  tetrads and the coordinate differentials 
 
\begin{equation}
\Theta^{(\alpha)}=g^{(\alpha)} _{\mu}dx^{\mu}.
\label{eq:6}
\end{equation}

We can rewrite equation (\ref{eq:5}) as tetrad componets

\begin{equation} 
\Theta^{(0)}=\frac{1}{2}Fdt+Ldr; \quad \Theta^{(1)}=dt;
\label{eq:7}
\end{equation}

\begin{equation} 
\Theta^{(2)}={\frac{r}{\sqrt{2}}(d\theta-i\;sin\theta \;d\varphi)}; \quad \Theta^{(3)}={\frac{r}{\sqrt{2}}(d\theta+i\;sin\theta\;d\varphi)}.
\label{eq:8}
\end{equation}
 
Then a square of space-time interval $ds^2$ will be written as 

\begin{equation} 
ds^{2}= g_{(\alpha)(\beta)}\Theta^{(\alpha)}\Theta^{(\beta)},
\label{eq:9}
\end{equation}

\noindent 
where $g_{(\alpha)(\beta)}$ is a tetrad metric, which in the matrix form can be written as 

\begin{center}
 \begin{equation}
 g_{(\alpha)(\beta)}=g^{(\alpha)(\beta)}=\left(\matrix{
 0 & 1 & 0 & 0\cr
 1 & 0 & 0 & 0\cr
 0 & 0 & 0 & -1\cr
 0 & 0 & -1 & 0\cr}\right).
\label{eq:10}
    \end{equation}
  \end{center}

The first equations of the Cartan structure are 
 
\begin{equation} 
{\bf d}\,\Theta^{(\alpha)}=-\omega^{(\alpha)}_{\;\;\;\;(\beta)}\wedge\Theta^{(\beta)}, 
\label{eq:11}
\end{equation}
 
\noindent 
where ${\bf d}$ is an exterior differential and a wedge operation $\wedge$ marks an exterior product.
The equations of structure allow to find non-zero 1-forms of a connection 

\begin{equation} 
 \omega_{(1)(0)}=\frac{1}{2}F'L^{-1}\Theta^{(1)}; \quad \omega_{(0)(2)}=(Lr)^{-1}\Theta^{(3)};
\label{eq:12}
\end{equation}

\begin{equation} 
\omega_{(0)(3)}=(Lr)^{-1}\Theta^{(2)}; \quad \omega_{(1)(2)}=-\frac{1}{2}F(Lr)^{-1}\Theta^{(3)}; 
\label{eq:13}
\end{equation}

\begin{equation} 
\omega_{(1)(3)}=-\frac{1}{2}F(Lr)^{-1}\Theta^{(2)}; \quad \omega_{(3)(2)}=\frac{1}{r\sqrt{2}}cot\theta(\Theta^{(2)}-\Theta^{(3)}),
\label{eq:14}
\end{equation}

\noindent 
a derivative with respect to radial variable $\;r$ is marked by a prime.

In general, we have for an arbitrary tetrad metric a true relationship 
${\bf d}\,g_{(\alpha)(\beta)}=\omega_{(\alpha)(\beta)}+\omega_{(\beta)(\alpha)},$
the left part of which equals zero under an investigation in our case because the tetrad metric components (\ref{eq:10}) are constants, i.e. the 1-forms of connection has an antisymmetric attribute

\begin{equation} 
\omega_{(\alpha)(\beta)} = -\omega_{(\beta)(\alpha)},
\label{eq:15}
\end{equation}

\noindent
This attribute allows to decrease a number of the independent 1-forms of connection. 

Further we make use of the second equations of the Cartan structure 

\begin{equation} 
\Omega^{(\alpha)} _{\quad(\beta)}=\frac{1}{2}R^{(\alpha)} _{\quad(\beta)(\gamma)(\delta)}\Theta^{(\gamma)}\wedge\Theta^{(\delta)}=
 {\bf d}\,\omega^{(\alpha)} _{\quad(\beta)}+\omega^{(\alpha)} _{\quad(\sigma)}\wedge\omega^{(\sigma)} _{\quad(\beta)},
\label{eq:16}
\end{equation}

\noindent 
where $\Omega_{(\alpha)(\beta)}=-\Omega_{(\alpha)(\beta)}$ is a 2-form of a curvature; $\;R^{(\alpha)}_{\quad(\beta)(\gamma)(\delta)}$ is Riemann's tensor of the curvature in the tetrad notations. 
Non-zero components of which can be found from (\ref{eq:16}) and equals to  

\begin{equation} 
R_{(1)(0)(1)(0)}=-\frac{F^{\prime\prime}}{2L^{2}}+\frac{F^{\prime}L^{\prime}}{2L^{3}};\quad R_{(1)(2)(1)(2)}=-\frac{F^{2}L^{\prime}}{4rL^{3}};
\label{eq:17}
\end{equation}

\begin{equation} 
R_{(0)(2)(0)(3)}=-\frac{L^{\prime}}{rL^{3}}; \quad R_{(0)(2)(1)(3)}=\frac{FL^{\prime}}{2rL^{3}}-\frac{F^{\prime}}{2rL^{2}}; 
\label{eq:18}
\end{equation}

\begin{equation} 
R_{(3)(2)(3)(2)}=\frac{1}{r^{2}}-\frac{F}{r^{2}L^{2}}; \quad R_{(0)(3)(1)(2)}=R_{(0)(2)(1)(3)}. 
\label{eq:19}
\end{equation}

According to the definition of Ricci's tensor ,
$R_{(\alpha)(\beta)}=R^{(\gamma)}_{\quad(\alpha)(\beta)(\gamma)}= g^{(\gamma)(\sigma)}R_{(\sigma)(\alpha)(\beta)(\gamma)},$
we will write his non-zero tetrad components:
 
\begin{equation} 
R_{(0)(0)}=-\frac{2}{rL^2}\frac{L^{\prime}}{L};\quad
R_{(0)(1)}=\frac{FL^{\prime}}{rL^3}-\frac{F^{\prime\prime}}{2L^2}+\frac{F^{\prime}L^{\prime}}{2L^3}-\frac{F^{\prime}}{rL^2}; 
\label{eq:20}
\end{equation}

\begin{equation} 
R_{(1)(1)}=-\frac{F^2}{2rL^2}\frac{L^{\prime}}{L};\quad
R_{(2)(3)}=-\frac{1}{r^2}\left(1-\frac{F}{L^2}-\frac{rF^{\prime}}{L^2}+\frac{rF}{L^2}\frac{L^{\prime}}{L}\right).
\label{eq:21}
\end{equation}

\section{ Einstein-Maxwell combined  equations \\
     for charged perfect fluid}

The Einstein equations in the tetrad description with a source in the form of the  
energy-momentum tensor (EMT) are

\begin{equation} 
G_{(\alpha)(\beta)}=R_{(\alpha)(\beta)}-\frac{1}{2}g_{(\alpha)(\beta)}R=-\varkappa T_{(\alpha)(\beta)}, 
\label{eq:22}
\end{equation}

\noindent 
where $G_{(\alpha)(\beta)}$ is the Einstein tensor;
$\;R_{(\alpha)(\beta)}$ is the Ricci tensor; $\;R =R^{(\alpha)}_{\quad(\alpha)}$ is the scalar curvature;
$\varkappa=8\pi$ is Einstein's gravitational constant in a chosen system of units. 
Resulting EMT of the matter $\; T_{(\alpha)(\beta)}$ is taken as direct sum of EMT of the perfect Pascal 
neutral fluid and EMT of the electromagnetic field

\begin{equation} 
T_{(\alpha)(\beta)}=T_{(\alpha)(\beta)}^{fluid}+T_{(\alpha)(\beta)}^{el-mag},
\label{eq:23}
\end{equation}

\noindent 
where 

\begin{equation} 
T_{(\alpha)(\beta)}^{fluid}=(\mu+p)u_{(\alpha)}u_{(\beta)}-pg_{(\alpha)(\beta)}\equiv\mu
u_{(\alpha)}u_{(\beta)}+pb_{(\alpha)(\beta)};
\label{eq:24}
\end{equation}

\begin{equation} 
T_{(\alpha)(\beta)}^{el-mag}=\frac{1}{4\pi}\left(-F_{(\alpha)(\sigma)}F_{(\beta)}^{(\sigma)}+\frac{1}{4}g_{(\alpha)(\beta)}F_{(\sigma)(\tau)}F^{(\sigma)(\tau)}\right);
\label{eq:25}
\end{equation}

\noindent 
$\mu(r)$ is a mass-energy density; $p(r)$ is a pressure of perfect Pascal fluid;
$u_{(\alpha)}=g_{(\alpha)\mu}\displaystyle\frac{dx^{\mu}}{ds}$ is a 4-velocity in tetrad notations;  $b_{(\alpha)(\beta)}=u_{(\alpha)}u_{(\beta)}-g_{(\alpha)(\beta)}$
is a 3-projector on a  spacelike hypersurface (or 3-metric) and $b_{(\alpha)(\beta)}u^{(\alpha)}=0;$ $F_{(\alpha)(\beta)}$ is the tensor of electromagnetic field with $F_{(\alpha)(\beta)}=-F_{(\alpha)(\beta)};$ all functions here depend on a radial variable only.

We will rewrite Einstein's equations (\ref{eq:22}) as 

\begin{equation} 
R_{(\alpha)(\beta)}=-\varkappa\left(T_{(\alpha)(\beta)}-\frac{1}{2}g_{(\alpha)(\beta)}T\right),
\label{eq:26}
\end{equation}

\noindent
using a connection between the scalar curvature and the trace of EMT, 
$R=\varkappa T.$

Then the gravitational combined equations for a spherical symmetry can be written as four equations 
for the dimensionless radial variable $x=r/R_0,$ which varies from zero to unit ($R_0$ is the exterior ball radius)

\begin{equation} 
\displaystyle\frac{2}{xL^2}(\ln{L})^{\prime}=\chi T_{(0)(0)}; 
\label{eq:27}
\end{equation}

\begin{equation} 
\displaystyle\frac{F^2}{2xL^2}(\ln{L})^{\prime}=\chi T_{(1)(1)};
\label{eq:28}
\end{equation}

\begin{equation} 
\displaystyle\frac{F}{xL^2}(\ln{L})^{\prime}-\frac{1}{2L^2}\left(F^{\prime\prime}+\frac{2}{x}F^{\prime}-F^{\prime}(\ln{L})^{\prime}\right)=-\chi\left(T_{(0)(1)}-\frac{1}{2}T\right); 
\label{eq:29}
\end{equation}

\begin{equation} 
\displaystyle\frac{1}{x^2}\left(-1+\frac{F}{L^2}+\frac{xF^{\prime}}{L^2}-x\frac{F}{L^2}(\ln{L})^{\prime}\right)=-\chi\left(T_{(2)(3)}+\frac{1}{2}T\right),
\label{eq:30}
\end{equation}

\noindent 
where all derivatives now are taken with respect to the variable $x\;$ and 
$\;\displaystyle\frac{\partial}{\partial r} = 
\displaystyle\frac{\partial}{R_0 \partial x} ,\;$ and a new constant is 
$\;\chi=\varkappa R^{2}_0 =  8\pi R^{2}_0$.

After that we will complete our system of equations (\ref{eq:27})-(\ref{eq:30}) with Maxwell's equations. The second pair of the Maxwell equations can be written in the framework of General Relativity as  
 
\begin{equation} 
F^{\mu\nu}_{\quad ;\nu}=\frac{1}{\sqrt{-g}}(\sqrt{-g}F^{\mu\nu})_{,\nu}= -4\pi\;j^{\mu},
\label{eq:31}
\end{equation}

\noindent 
where $j^{\mu}$ is a density of an electric current, and a point with a comma marks a covariant derivation. 
The tensor of the electromagnetic field is set as usual over a alternation of 4-potential derivations 
$A_{\mu}$ как $F_{\mu\nu}=A_{\nu,\mu}-A_{\mu,\nu}.$ Only one component of the tensor of the electromagnetic field here will be non-zero $F_{01}=-A_{0,1}=-R_0\;A_{0}^{\prime}$. Then the Maxwell combined equations will be transformed to the one equation

\begin{equation} 
\frac{1}{x^{2}L}(x^{2}LF^{01})^{\prime}=-4\pi\,R_0\, j^{0}. 
\label{eq:32}
\end{equation}

Now we will take a comoving reference of frame in which the 4-velocity has the following components 
in a static case 

\begin{equation} 
u^{\mu}=\frac{\delta^{\mu}_{0}}{\sqrt{g_{00}}}\equiv\frac{\delta^{\mu}_{0}}{\sqrt{F(r)}};\quad
u_{\mu}=\frac{g_{\mu0}}{\sqrt{g_{00}}}\equiv\frac{g_{\mu0}}{\sqrt{F(r)}}.
\label{eq:33}
\end{equation}

\noindent 
The physical observed magnitudes will be rewritten respectively as 

\begin{equation} 
\mu_{phys}=T^{fluid}_{\mu\nu}u^{\mu}u^{\nu}=\mu(x); \quad \rho_{phys}\equiv\rho(x)=j^{\mu}u_{\mu}=j^{0}\sqrt{g_{00}}\equiv j^{0}\sqrt{F(x)};
\label{eq:34}
\end{equation}

\begin{equation} 
E^{phys}_{\nu}=-F_{\nu\mu}u^{\mu}=-F_{\nu\mu}\frac{\delta^{\mu}_{0}}{\sqrt{g_{00}}}=\frac{F_{0\nu}}{\sqrt{F(x)}}; \quad
E_{phys}\equiv E_{1}=\frac{F_{01}}{\sqrt{g_{00}}}=\frac{E}{\sqrt{F(x)}}; 
\label{eq:35}
\end{equation}

\begin{equation} 
W_{el}=T^{el-mag}_{\mu\nu}u^{\mu}u^{\nu}=\frac{E^{2}}{8\pi L^{2}},
\label{eq:36}
\end{equation}

\noindent 
where $\mu_{phys}$ is a physical observed mass-energy density; $\;\rho_{phys}$ is a physical observed density 
of electrical charge; $\;E^{phys}_{\nu}$ is physical observed 3-vector of  electric field strength; $\;E_{phys}\equiv E_{1}$ is a radial component of physical observed of electric field strength; 
$\;W_{el}$ is a physical observed of electric field energy density. 

Thus the Maxwell equations (\ref{eq:32}) are rewritten in a form for the further using

\begin{equation} 
\left(\frac{x^{2}E}{L}\right)^{\prime}=4\pi\rho(x)R_0 \frac{x^2}{\sqrt{\varepsilon}} ,
\label{eq:37}
\end{equation}

\noindent
where we set a new function $\varepsilon = \displaystyle\frac{F}{L^2}.$

Total EMT can be written in the tetrad components as

\begin{equation} 
T_{(0)(0)}=\frac{1}{F}(\mu+p); \quad T_{(1)(1)}=\frac{1}{4}F(\mu+p); 
\label{eq:38}
\end{equation}

\begin{equation} 
T_{(0)(1)}=\frac{1}{2}(\mu-p)+W_{el}; \quad T_{(2)(3)}=T_{(3)(2)}=p+W_{el}. 
\label{eq:39}
\end{equation}

\section{Transformation of Einstein's equations}

\indent
 
The system of equations after a substitution of the expressions (\ref{eq:38})-(\ref{eq:39}) into the 
right part of gravitational equations (\ref{eq:27})-(\ref{eq:30}) are reduced to

\begin{equation} 
\displaystyle\frac{\varepsilon}{x}(\ln{L})^{\prime}= \displaystyle\frac{\chi}{2}(\mu+p); 
\label{eq:40}
\end{equation}

\begin{equation} 
\displaystyle\frac{\varepsilon}{x}(\ln{L})^{\prime}-\frac{\varepsilon}{2}\left(\frac{F^{\prime\prime}}{F}+\frac{2}{x}(\ln{F})^{\prime}-{(\ln{F})^{\prime}}(\ln{L})^{\prime}\right)=-\chi\left(p +W_{el}\right); 
\label{eq:41}
\end{equation}

\begin{equation} 
-\displaystyle\frac{1}{x^2}(1-\varepsilon)+\displaystyle\frac{\varepsilon}{x}\left(\ln{\frac{F}{L}}\right)^{\prime}=-\chi\left(\frac{1}{2}(\mu -p)+W_{el}\right).
\label{eq:42}
\end{equation}

\noindent 
After that we exclude the pressure and the mass-energy density and get a linear differential equation with variable coefficients for function  $G(x)$ 

\begin{equation} 
G^{\prime \prime}+f(x)G^{\prime}+g(x)G=0,
\label{eq:43}
\end{equation}

\noindent
where $G=\sqrt{F},\;$ $f(x)=(\ln{\varphi})^{\prime}$, $\varphi(x)=\sqrt{\varepsilon}/x,$
and a coefficient $g(x)$ equals to

\begin{equation} 
g(x)=\frac{2(1-\varepsilon)+x\varepsilon^{\prime}}{2x^2\varepsilon}-\frac{2\chi}{\varepsilon}W_{el}.
\label{eq:44}
\end{equation}

\noindent
An electromagnetic field influence is appeared  directly through the functions  $\varepsilon(x)\;$ and $\; W_{el}(x).$ 

The definition of a new variable $\zeta = \zeta(x)$, according to an expression 

\begin{equation} 
d{\zeta} = \displaystyle\frac{x dx}{\sqrt{\varepsilon(x)}},
\label{eq:45}
\end{equation}

\noindent
transforms (\ref{eq:43}) into an equation of a nonlinear spatial oscillator 

\begin{equation} 
G^{\prime \prime}_{\zeta \zeta}+\Omega^2(\zeta(x))G=0,
\label{eq:46}
\end{equation}
 
\noindent
with respect to variable $\zeta.\;$ The square of ``frequency'' $\;\Omega^2$ easier to use in the form 

\begin{equation} 
\Omega^2 = - \displaystyle\frac{d}{dy}\left(\frac{\Phi}{y}\right)- \displaystyle\frac{2\chi}{y} W_{el}, 
\label{eq:47}
\end{equation}

\noindent
because we have a difficulties with an integration of the expression (\ref{eq:45}) in the elementary functions. In the equation (\ref{eq:47}) $y = x^2,\;$ and a function $\Phi$ is an analog of Newton's gravitational potential of an interior region of a charged fluid ball. The function $\Phi$ can be found from gravitational equations through the function $\varepsilon$ as  

\begin{equation} 
\Phi=1-\varepsilon=\frac{\chi}{x}\int(\mu(x)+W_{el}(x))x^2 dx = \frac{\chi}{2\sqrt{y}}\int(\mu(y)+W_{el}(y))\sqrt{y} dy .
\label{eq:48}
\end{equation}

Also an expression for pressure can be found easily from the gravitational combined equations (\ref{eq:40})-(\ref{eq:42})

\begin{equation} 
\chi p = \chi W_{el}-\displaystyle{\frac{\Phi}{x^2}+\frac{1}{x}}(1-\Phi)(\ln F)^{\prime}.
\label{eq:49}
\end{equation}

\section{Boundary conditions}

\indent 
The interior solution must be smoothly sewed together on the ball surface
 ($r = R_0$) with the exterior solution of Reissner-Nordstr$\ddot{o}$m with metric (\ref{eq:1}). The functions $g_{00}$ and $g_{01}$ are written in an exterior space-time as
 
\begin{equation}
g_{00} = F_{R-N}(r)= 1-\displaystyle{\frac{2m}{r} + \frac{Q^2}{r^2}}; \quad g_{01} = L_{R-N} = 1,
\label{eq:50}
\end{equation}

\noindent
where $m$ and $Q$ are an integral mass and an integral electric charge respectively which are calculated by observer on a spatial infinity.

If we set a parameter of compactness $\eta = \displaystyle\frac{2m}{R_0}$, which describes a level of gravitational compression of a star then we will have the expressions for the functions $F(x)=G^2(x)$ and $\varepsilon(x)$ on a surface of ball 

\begin{equation}
F_{(x=1)}=G^2_{(x=1)}= \varepsilon_{(x=1)}=1-\eta+\displaystyle\frac{Q^{2}}{R_0^{2}} = 1-\eta^{*},
\label{eq:51}
\end{equation}
  
\noindent
where $\eta^{*}= \eta - \displaystyle\frac{Q^{2}}{R_0^{2}}$ is an effective compactness. Here we must remark that $\eta^{*} \leq \eta.$  

The function $L$ automatically is continuous on a boundary $r = R_0$ and the function $\Phi_{(x=1)} = \eta^{*}.$

The continuity requirement of a function $F(x)$ on the ball surface leads to

\begin{equation}
(F_{(x=1)})^{\prime}= 2(G_{(x=1)})^{\prime} G_{(x=1)} = 2\eta^{*}-\eta.
\label{eq:52}
\end{equation}

\noindent
Therefore, we have

\begin{equation}
(G_{(x=1)})^{\prime}= \displaystyle\frac{2\eta^{*}-\eta}{2\sqrt{1-\eta^{*}}}.
\label{eq:53}
\end{equation}

Now we will return to the expression for pressure (\ref{eq:49}) and will require of pressure's absence on the ball surface because there are vacuum in an exterior region. The boundary conditions were found for the physical functions and lead to an equation for parameters 

\begin{equation}
\chi W_{el} = \eta -\eta^{*} = \displaystyle\frac{Q^2}{R_0^2}.
\label{eq:54}
\end{equation}

\noindent
This expression is transformed in an identity, when the electric charges are absent
($\eta = \eta^{*}$).

\section{Solution of Einstein-Maxwell equations}
\indent 

Unlike paper [1], where a mass density of a neutral fluid has the parabolic distribution law in an interior region of the ball (Fig.1) 

\begin{equation} 
\mu(x)=\mu_{0}(1-b\,x^2),
\label{eq:55}
\end{equation}

\noindent
we will consider the behaviour of a system for a more general mass density distribution of the neutral substance (Fig.2)

\begin{equation} 
\mu(x)=\mu_{0}(1-b\,x^2)^{3}. 
\label{eq:56}
\end{equation}


\begin{figure}[hbt]
\begin{minipage}[t]{0.47 \linewidth}
\centering {\includegraphics [bb = 0 0 6.9cm 7.0cm] {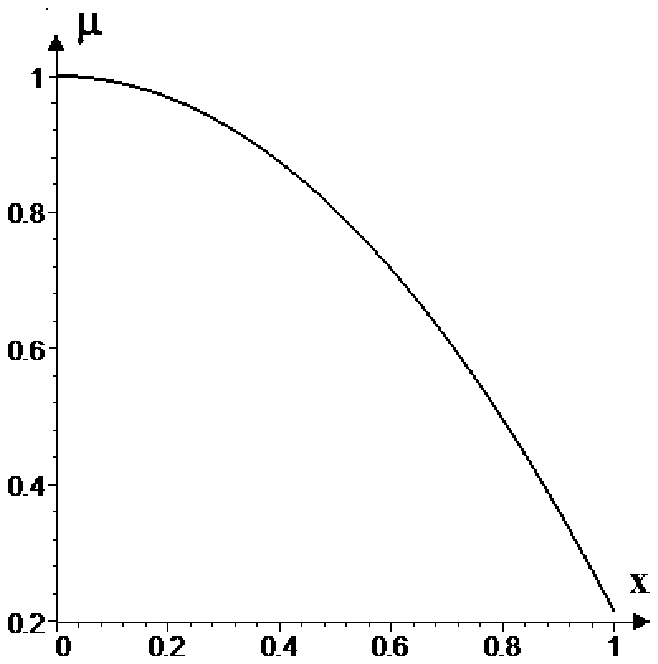}}
\caption{Behaviour of parabolic 
mass den- \\sity distribution as function of dimension-\\
less 
radial variable 
for parameter $b=63/80.$}
\end{minipage}
\hfill \begin{minipage}[t]{0.47 \linewidth}
\centering {\includegraphics [bb = 0 0 6.9cm 7.0cm] {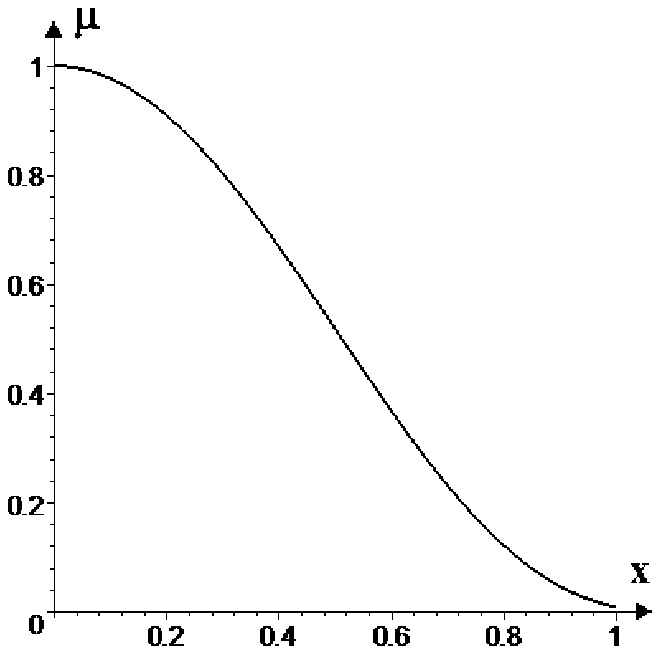}}
\caption{Behaviour of mass density distri-\\
bution as function of dimensionless radial \\
 variable 
on base of parabolic mass density \\
behaviour for parameter $b=63/80.$}
\end{minipage}
\end{figure}


An expression of the electrical field energy density $W_{el}$ can be generalized if we will require when the gravitational field disappears, i.e. $G_{N}\rightarrow 0$ , then  the expression for $W_{el}$ will equal to the electrical field energy density for the homogeneous charged ball ($\rho\rightarrow\rho_{0}=const$) in a flat space-time

\begin{equation} 
W_{el}=\displaystyle\frac{\lambda_0\, x^2 }{8 \pi}, 
\label{eq:57}
\end{equation}

\noindent
where $\lambda_0 = const.$  

Therefore we will determine the electrical field energy density unlike the article [1] as 

\begin{equation} 
W_{el}=\frac{1}{8\pi}\lambda^2(x)\, x^2,
\label{eq:58}
\end{equation}

\noindent 
and we will generalize also the expression of the electrical charged density so that a new expression of the electrical charged density with $a = 0$ will be equal to the similar expression in [1]

\begin{equation} 
\rho=\rho^{*}\sqrt{\varepsilon(x)}=\rho_{0}(1-ax^2)\sqrt{\varepsilon(x)}. 
\label{eq:59}
\end{equation}

Then we find directly from the Maxwell equation (\ref{eq:37}) 

\begin{equation} 
\lambda(x)=\frac{4\pi\rho_{0}R_0}{3}\left(1-\frac{3a}{5}x^2\right).
\label{eq:60}
\end{equation}

We find the final expression of the electrical charged density after substitution (\ref{eq:60}) into (\ref{eq:58})

\begin{equation} 
\chi W_{el}=\left(\frac{4\pi
R_0^2}{3}\right)^2\rho_{0}^2\left(1-\frac{3a}{5}x^2\right)^2x^2.
\label{eq:61}
\end{equation}

We will find also a restriction on a parameter region $a:$ $0 < a \leq 5/3.$ from the expression (\ref{eq:54}) for the pressure value on the ball surface and under using equation (\ref{eq:61}). However a non-negative behaviour of the electric charge density (\ref{eq:59}) restricts else more the parameter region: $0 < a \leq 1.$ We must remark that a sign of an electric charge is set with a parameter $\rho_0.$ 

Furthermore, the boundary conditions 
$\rho(x=0)\equiv\rho(0)=\rho_{0}\sqrt{\varepsilon(0)}$; $\rho(x=1)\equiv\rho(1)=\rho_{0}(1-a)\sqrt{\varepsilon(1)}$ 
lead us to the expression

\begin{equation} 
a=1-\frac{\rho(1)}{\rho(0)}\frac{\sqrt{\varepsilon(0)}}{\sqrt{\varepsilon(1)}}.
\label{eq:62}
\end{equation}

If we connect parameters $a$ and $b$ in this case with the connection $a=80/63b$, then we will have the constancy condition of a squared ``frequency'' $\;\Omega_0^{2}=const.$

Further we can rewrite (\ref{eq:46}) as the equation for the harmonic spatial oscillator.

\begin{equation} 
G^{\prime \prime}_{\zeta \zeta}+\Omega_0^2\,G=0,
\label{eq:63}
\end{equation}

\noindent 
where after the using connection between parameters $a$ and $b$ an expression for $\;\Omega_0^{2}$ can be written as 

\begin{equation} 
\Omega_0^{2} = \displaystyle\frac {\chi}{60^2}(1701\mu_0 a - 1760 \pi \rho_0^2 R_0^2).
\label{eq:64}
\end{equation}

A general solution can be directly written as a harmonic oscillating function for such an equation under condition that the squared "frequency"$\Omega_0^{2}$ is positive 

\begin{equation} 
G(\zeta(x))=G_{0}cos(\Omega_{0}\cdot\zeta(x)+\alpha_0), 
\label{eq:65}
\end{equation}

\noindent
where $\alpha_0$ is a phase shift.

The total expression of metric function $g_{00} = F = G^2$ will be 

\begin{equation} 
F(x)=G_{0}^2 \,cos^2(\Omega_{0}\cdot\zeta(x)+\alpha_0),
\label{eq:66}
\end{equation}

\noindent
and a function $g_{01} = L$ is found lightly from the expression $L = \displaystyle\sqrt{\frac{F}{\varepsilon}}.$

We can find the constants $G_0$ and  $\alpha_0$ from the boundary conditions, expressing in terms of  known parameters

\begin{equation} 
G_0 = \left( 1- \eta^{*}+ \displaystyle\frac{2 \eta^{*}-\eta}{2 \Omega_0}\right)^{1/2};
\label{eq:67}
\end{equation}

\begin{equation} 
\tan(\Omega_0\,\zeta(x=1)+\alpha_0)= 
-\displaystyle\frac{2 \eta^{*}-\eta}{2 \Omega_0 \sqrt{1-\eta^{*}}}.
\label{eq:68}
\end{equation}

In the summary it is necessary to remark that in the article a reduction method of the Einstein-Maxwell equations for a static spherical case to the equation of a nonlinear spatial oscillator is demonstrated. These equations were written in Bondi's radiation coordinates and with a source in the perfect charged Pascal fluid form. An exact solution in the framework General Relativity as the solution of an equation for a spatial harmonic oscillator and concrete relationship between the physical parameters which were included in the researching problem is found. Also an assumption on the mass-energy density behaviour and the electric field energy density in an interior region of the charged fluid gravitational ball was introduced.
 This solution is a generalization of an earlier finding solution in [1].

\end{document}